# High-speed measurement of firearm primer blast waves

Michael Courtney, Joshua Daviscourt, Jonathan Eng
U.S. Air Force Academy, 2354 Fairchild Drive, USAF Academy, CO, 80840
Michael.Courtney@usafa.edu

Amy Courtney
Force Protection Industries, Inc., 9801 Highway 78, Ladson, SC 29456
amy_courtney@post.harvard.edu

**Abstract:** This article describes a method and results for direct high-speed measurements of firearm primer blast waves employing a high-speed pressure transducer located at the muzzle to record the blast pressure wave produced by primer ignition. Key findings are: 1) Most of the lead styphnate based primer models tested show 5.2-11.3% standard deviation in the magnitudes of their peak pressure. 2) In contrast, lead-free diazodinitrophenol (DDNP) based primers had standard deviations of the peak blast pressure of 8.2-25.0%. 3) Combined with smaller blast waves, these large variations in peak blast pressure of DDNP-based primers led to delayed ignition and failure to fire in brief field tests.

## I. Introduction

Over the years various surrogates have been used to quantify and compare performance of rifle primers including measuring velocity and standard deviation when the primer alone propelled a projectile from a gun barrel,(1) measuring velocity, pressure, and standard deviation produced by a given primer in combination with a given powder charge and bullet,(2)(3) and measuring the size of the visible primer flash in photographs.(2)(3) This article presents a method and results for direct high-speed measurements of rifle primer blast waves employing a high-speed pressure transducer located at the muzzle to record the blast pressure wave produced by primer detonation and by showing that mass sorting produces a smaller deviation in peak primer pressures.

It is commonly reported that choosing the least powerful primer that can reliably ignite a powder charge often produces the smallest standard deviations in muzzle velocity, thus the smallest vertical dispersions at long range. Two causal hypotheses have emerged for this observation. Lapua's published brochure on the .308 Winchester Palma Case featuring a small rifle primer pocket describes the idea that small rifle primers themselves simply exhibit less variations. The other hypothesis is (in the words of German Salazar), "accuracy is more easily found when the influence of the primer on the overall pressure of the load is minimized."(3) The data presented here is inconclusive regarding which hypothesis is more correct; however, the measurement method presented could be used, together with mass sorting and measurement of velocity standard deviations to determine which hypothesis is better supported in a given cartridge and load.

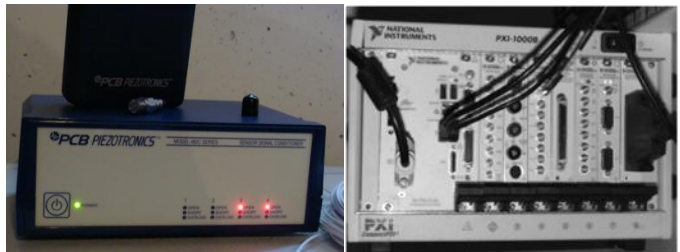

*Figure 1: Left: High-speed pressure transducer (on top) and signal conditioning unit. Right: Fast waveform digitizer in PXI system.*

## II. Method

Rifle primers work by the impact detonation of high-explosive compounds (usually a combination of lead styphnate and lead azide in modern primers), which then ignites the propellant charge. The measurement method is simple: a firearm loaded with a primed cartridge case without any gunpowder or projectile has all the essential elements of an explosive driven shock tube whose shock wave is emitted from the muzzle after the primer is detonated by the firing pin. The blast wave measured at the muzzle depends on the strength of the primer without the confounding factors (bore friction, neck tension, powder charge, bullet bearing surface, cartridge case variations, etc.) that affect other methods of inferring primer strength and consistency.

# High-speed measurement of firearm primer blast waves

Here, a Remington 700 ADL chambered in .308 Winchester with a 22" barrel is used for the test platform. Tests on large rifle primers employ R-P brass with the pockets uniformed with the Redding tool, and the flash hole deburred with a handheld center drill from the outside and an oversized drill bit from the inside. Primers are loaded into the case with an RCBS Rockchucker reloading press. Tests on small rifle primers employ the Lapua Palma case featuring a small rifle primer pocket prepared in the same manner.

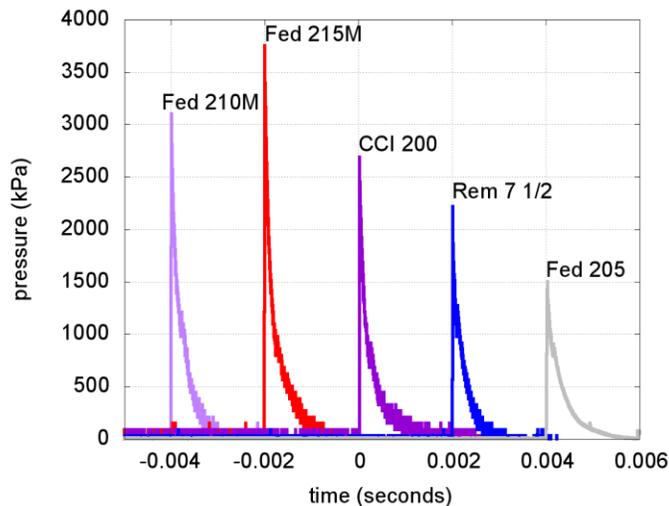

*Figure 2: Typical blast pressure waveforms measured for several rifle primer types. The detonation times have been shifted in 0.002 second (2 millisecond) increments to better visualize and compare waveform shapes.*

The blast pressure measurements presented here result from using high-speed pressure transducers (PCB 102B and PCB 102B15) specifically designed for measuring the very fast pressure transients associated with explosive detonations and other shock waves. The pressure transducer is placed coaxially with the rifle barrel and directly facing the muzzle with no separation between the end of the barrel and pressure transducer. A cable connects the transducer to a signal conditioning unit (PCB 842C) which produces a calibrated voltage output which is then digitized with a National Instruments PXI-5105 fast analog to digital converter operating at a rate of 1 million samples per second. The voltage waveform is saved as a file for later conversion to pressure using the calibration certificate provided by the manufacturer with each pressure sensor. A high-speed pressure transducer, signal conditioning unit, and fast waveform digitizer are shown in Figure 1.

### III. Results

Figure 2 shows blast pressure waveforms for several rifle primer types. The waveforms are combined on a single graph to facilitate comparison. Dozens of these waveforms were measured for the study reported here, but rather than show all the graphs, it is more revealing to characterize the waveform shapes with their key parameters and then discuss the average and standard deviation because these best characterize primer strength and consistency.

Simple blast waves are usually characterized by peak overpressure, duration, and impulse (the area under the curve of pressure vs. time). Since the durations and basic shapes are all nearly the same for all the pressure waveforms, the impulse is nearly proportional to the peak pressure, and the peak pressure is the main distinguishing characteristic of the blast wave. Therefore, we will focus on the average peak magnitude and the standard deviation of peak magnitudes for each primer type.

| Primer | Diameter (mm) | Peak Pressure (kPa) | SD (kPa) | SD (%) |
|---|---|---|---|---|
| Fed 210M | 5.33 | 2908 | 223 | 7.7% |
| Fed 215M | 5.33 | 3811 | 192 | 5.0% |
| CCI 200 | 5.33 | 2561 | 270 | 10.7% |
| CCI 250 | 5.33 | 3587 | 404 | 11.3% |
| **DDNP KVB-7E** | **5.33** | **1186** | **296** | **25.0%** |
| Rem 7 ½ | 4.45 | 2303 | 186 | 8.1% |
| Fed 205 | 4.45 | 1469 | 103 | 7.1% |
| CCI 450 | 4.45 | 1602 | 104 | 6.5% |
| Fed 205M | 4.45 | 1434 | 103 | 7.2% |
| **DDNP KVB-9E** | **4.45** | **1331** | **109** | **8.2%** |

*Table 1: Peak pressure averages and standard deviations from the mean (SD) with a sample size of 10.*

Table 1 shows average peak pressures along with standard deviations from the mean for a selection of both large and small rifle primers. As expected, large rifle primers usually produce stronger blast waves than small primers, and "magnum" rifle primers

# High-speed measurement of firearm primer blast waves

(Fed215M, CCI250, CCI450) produce stronger blast waves than non-magnum primers of the same size. There are significant differences in the standard deviations observed for different primer types, and it is notable that so-called "Match" primers are not always more consistent than non-match primers. The two DDNP-based primers have the largest variations in their size with the KVB-7E having very large variations. Figure 3 compares graphs of the largest and smallest blast waves of the two DDNP-based primers with two lead styphnate based primers.

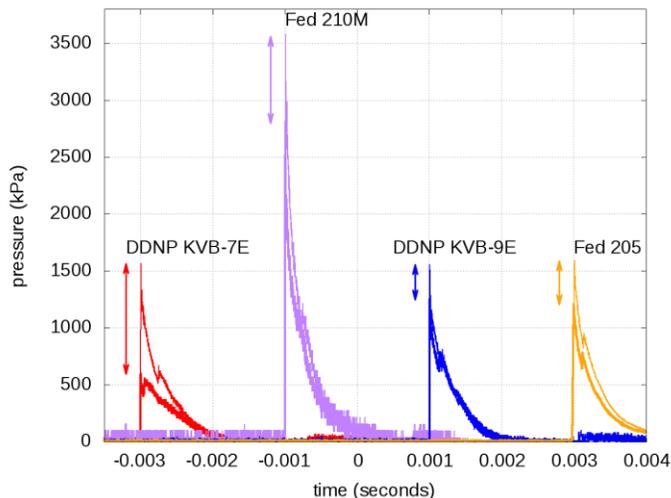

Figure 3: Largest and smallest blast pressure waves are shown for two DDNP-based (KVB-7E and KVB-9E) and two lead based (Fed 210M and Fed 205) primers.

Since it is of interest to know how much these blast pressure differences impact field performance, some brief field testing was conducted comparing 10 shots with the DDNP-based KVB-7E rifle primer with 10 shots of the lead styphnate based Fed 210M in each of two otherwise identical loads: 1) a 30-06 load using 51.0 grains of H414 (a ball powder) in Remington brass with a 220 grain Sierra MatchKing bullet and 2) a 7.62x51mm NATO load using 46.0 grains of Varget (an extruded powder) using Remington brass with a Berger 155.5 grain Fullbore boat tail bullet. Both tests were conducted with Remington 700 rifles in HS Precision stocks. The most obvious difference between the lead based and DDNP-based primers was a perceptible delay between firing pin strike and ignition in 15 of 19 shots with the DDNP-based primers (and one misfire); in contrast, there were no misfires or perceptible delays in ignition with the lead based primer. (In fact, in over one thousand rounds using lead based primers in these two rifles, the authors have never observed a perceptible delay in firing nor a misfire.) Figure 4 shows the primer which failed to ignite the powder charge and resulted in a misfire.

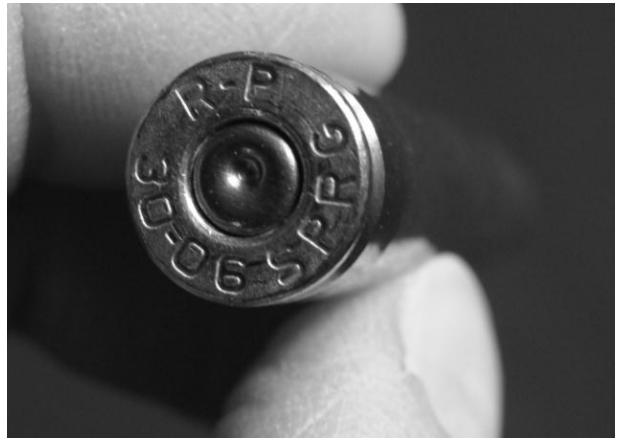

Figure 4: DDNP-based KVB-7E primer which produced a misfire in a 30-06 test load. The crater suggests the misfire was not due to a light primer strike. The other nine 30-06 test loads with this primer demonstrated a perceptible delay in ignition.

Excluding the misfire, the average velocity of the 30-06 load was lower (703 m/s) for the DDNP-based primer compared with the lead based primer (727 m/s). The standard deviation in muzzle velocities was comparable for the DDNP-based primers (4.8 m/s) and for the lead based primers (4.4 m/s) in the 30-06 load, and the average 5 shot group size (extreme spread measured at 200 m) was 2.5 minutes of angle (MOA) for the DDNP-based primers and 2.4 MOA for the lead based primers.

In the 7.62x51mm NATO load, both primers produced an average muzzle velocity of 823 m/s with the DDNP-based primer giving a smaller standard deviation (2.8 m/s) than the more powerful lead styphnate based primer (6.5 m/s). This agrees with the hypothesis (3) that having a primer that is not more powerful than needed to reliably ignite the powder charge produces more consistent muzzle velocities than a more powerful primer. The delay in ignition in 6 of the 10 shots with the DDNP-based primer suggests that this primer is at the low end of strength needed to reliably ignite 46 grains of an extruded powder. This ignition delay is the most likely cause of the larger average group size (2.5 MOA) of the DDNP-based primers in the 7.62x51mm NATO load compared with the lead styphnate based primers (1.8 MOA) at 200 m.

### IV. Discussion

Key findings are: 1) Most of the lead styphnate based primer models tested show 5.0-11.3% standard deviation in the magnitudes of their peak pressure. 2) In contrast, lead-free DDNP-based primers had standard deviations of the peak blast pressure of 8.2-25.0%. 3) Combined with smaller blast waves, these large variations in peak blast pressure led to delayed ignition and failure to fire in brief field tests.

The history of primer technology is somewhat cyclical with several notable instances of new primer chemistry being introduced to better meet an environmental or gun maintenance concern with several decades passing before the new chemistry became reliable. In 2010, the Office of the Product Manager for Maneuver Ammunition Systems projected that green primer formulations for use in the U.S. military will be evaluated and candidates selected in FY 2011, and that ammunition with green primers will be at full production by the end of FY 2012.(6) At the turn of the 20$^{th}$ century, primer development was driven by the need for a non-corrosive formulation. In the following years, changes in primers used by the military were necessary due to lack of shelf-stability, which led to misfires. This was a reason the U.S. military moved from mercury fulminate-based primers prior to WWI to a formulation based on potassium chlorate, antimony trisulphide and sulphur. However, this formulation was associated with misfires and corrosion, forcing another change.(7)

The lesson of primer history is that care is needed to prevent another large scale move to new primer technology that will compromise field performance and produce unintended consequences. Since difficulty obtaining consistent field performance from lead-free rifle primers was observed in this study and has been noted by others,(8)(9) we recommend independent testing demonstrate the following characteristics before any DDNP-based primer is adopted for duty:

1. Peak blast wave magnitude and consistency comparable with lead based primers.
2. Misfire rates at or below those with lead based primers.
3. Shelf-life and long term stability comparable with lead based primers.
4. Muzzle velocity consistency and peak chamber pressure comparable with lead based primers.
5. Ignition delay times comparable with lead based primers.
6. Comparable accuracy with lead based primers in both machine rests and hand-held testing.

**Acknowledgements:**
The authors acknowledge and appreciate the use of test equipment from Force Protection Industries, Inc. We are also grateful to German Salazar for providing helpful suggestions and comments on the manuscript as well as resources and to Leo Ahearn of the Colorado Rifle Club for providing resources on short notice.